# Exploring Physics Teachers' Views on Physics Education Research: A Case of Science Scepticism?

Melissa Costan^a*, Kasim Costan^a, Anna Weißbach^a and Christoph Kulgemeyer^a

^aDepartment of science education research, University of Bremen, Bremen, Germany

*Melissa Costan is a PhD candidate in physics education at the University of Bremen. She works on science scepticism towards science education research. melissa3@uni-bremen.de

https://orcid.org/0009-0005-7881-7162

Kasim Costan is a PhD candidate in physics education at the University of Bremen. He works on the impact of science scepticism on physics teachers' professional development. kasim@uni-bremen.de

- https://orcid.org/0009-0007-8527-0040

Anna Weißbach has a PhD in physics education at the University of Bremen. Her work focuses on physics teachers' reflection skills. anna.weissbach@uni-bremen.de

- https://orcid.org/0000-0003-1183-971X

Christoph Kulgemeyer is a professor of physics education at the University of Bremen. He works on physics teachers' professional knowledge, instructional explanations, and reflection skills. His latest work focuses on how to learn effectively using physics explainer videos. kulgemeyer@physik.uni-bremen.de

- https://orcid.org/0000-0001-6659-8170

Funding details: none

# Exploring Physics Teachers' Views on Physics Education Research: A Case of Science Scepticism?

The gap between theory and practice is well-documented in educational research. Physics teachers' willingness to apply research findings in practice may be influenced by a sceptical attitude towards science education research. This study explores physics teachers' perspectives on science education research, with a particular focus on potential scepticism towards the discipline. A two-step mixed-methods approach was employed: (1) Interviews with a purposeful sample of 13 experienced physics teachers for a first exploration of attitudes towards physics education research, and (2) a quantitative survey of 174 physics teachers to examine, among other aspects, the previously observed attitudes in a larger sample and to identify teacher profiles using latent profile analysis. The interview study revealed both sceptical and non-sceptical attitudes towards physics education research, including some that fundamentally questioned its practical value. Based on the survey data and latent profile analysis, four distinct teacher profiles differing in their level of scepticism towards science education research were identified. While one profile is highly sceptical, the other three exhibit a mix of sceptical and supportive attitudes. Thus, physics teachers are not generally sceptical. However, the cooperation between research and practice is perceived as unproductive by most teachers.

**Introduction**

Most interestingly, the perspective of science teachers – especially those with extensive teaching experience – on science education research is largely unexplored. This research gap is surprising for at least two reasons: (1) the theory-practice gap, and (2) the assumed practical value of evidence-based teaching.

Regarding (1), the 'theory-practice gap' is a well-documented phenomenon in various educational research disciplines, including science education. Particularly novice teachers often perceive a disconnect between their academic teacher education and their daily practice, which can sometimes lead to feeling overwhelmed (Dicke et al., 2016). Research also suggests that teaching materials developed based on evidence are not always well-received by teachers (Breuer, 2021). Nägel et al. (2023) define science scepticism towards general educational research as scepticism regarding its practical value. This scepticism towards the scientific discipline of science education research may contribute to both phenomena of the theory-practice gap and the assumed practical value of evidence-based teaching: If individuals perceive little value in scientific evidence for practical teaching, this could contribute to a perceived gap between academic teacher education and teaching practice, as well as a lack of appreciation for teaching materials resulting from such research.

Regarding (2), science education research typically assumes the practical value of at least some of its findings. We would argue that this assumption is even fundamental to the entire scientific discipline and many researchers view evidence-based teaching as crucial for enhancing the overall quality of science education (e.g., Nelson & Campbell, 2017). However, if science teachers harbour scepticism regarding the value of science education research as a discipline, its methodologies, or its findings, this may affect their willingness to incorporate scientific insights from educational research into their teaching practices. Consequently, they may overemphasize the value of their personal experiences compared to generalizable scientific findings. In this regard, it is important to acknowledge that attitudes – including scepticism towards science education research – are known to directly influence actions: Attitudes serve to define tasks in ill-structured situations as those encountered in teaching practice to facilitate teachers' available attention (Nespor, 1987), and they play a crucial role in determining the extent to which professional knowledge is applied in practice (Renkl, 1996). Thus, it is rather surprising that there is limited research exploring how teachers perceive the evidence they should utilize and the discipline responsible for producing this evidence.

In the present paper, we aim to contribute to exploring physics teachers' perspectives on science education research, with a focus on uncovering potential scepticism towards physics education research among teachers. This study adopts a two-step mixed-methods approach: (1) an interview study involving a purposeful sample of 13 physics teachers, aimed at developing categories describing their attitudes towards physics education research and its practical value; and (2) a quantitative survey targeting a larger group of 174 physics teachers (convenience sample), using (among other things) the categories identified in study 1 to construct rating scales. The results of this study will be subjected to descriptive analyses, with a focus on identifying teacher profiles using latent profile analysis.

**Literature Review**

***The gap between theory and practice***
The demand for evidence-based practice assumes that educational research findings, like innovative teaching materials (e.g. Wilhelm et al., 2021) or studies on more general design principles like inquiry-based instruction (e.g. Vorholzer et al., 2022), are implemented into teaching practice (Schrader et al., 2020). However, research shows that the implementation of these findings into practice is only partially successful. For example, more often than not, innovative teaching materials do not establish in teaching practice, even if they are empirically tested and deemed successful (Altrichter & Wiesinger, 2004). This is referred to as the gap between theory and practice and has been investigated by many researchers (e.g., Broekkamp & van Hout-Wolters, 2007; Nägel et al., 2023; Vanderlinde & van Braak, 2010). The theory-practice gap is a complex phenomenon that is analysed and evaluated differently by researchers and educators (Broekkamp & van Hout-Wolters, 2007; Vanderlinde & van Braak, 2010). Compounding the issue is the fact that researchers and educators hold differing conceptions of educational research (Edwards et al., 2007). In the field of educational research, researchers often disagree on the understanding, nature, objectives, and methodologies of the discipline (Vanderlinde & van Braak, 2010). Educators, on the other hand, often demand the immediate applicability of research findings and criticize that they are too theoretical (Merzyn, 2004). For instance, research suggests that teachers encounter difficulties in analysing pedagogically relevant situations based on scientific evidence and making appropriate decisions accordingly (Stark et al., 2010). However, successful implementation requires teachers to engage with scientific evidence and apply it in their practice. As evident, considering different perspectives is essential for understanding the theory-practice gap, which, of course, cannot be fully addressed within a single study. In this study, we aim to examine the perspective of teachers towards science education research, as it is already well known that their attitudes towards educational research play a key role in the implementation of innovations, yet these attitudes have rarely been systematically studied (Gräsel & Parchmann, 2004; Schrader et al., 2020). Regarding educational research, there is some evidence that teachers' attitudes towards it play a crucial role in the implementation process: In the study by Vanderlinde and van Braak (2010), teachers argue that educational researchers do not ask questions relevant to practice and that research findings often are ambiguous or even contradictory. Schaik et al. (2018) identified teachers' attitudes towards research knowledge as a significant barrier to the utilization of academic knowledge. Similarly, Lysenko et al. (2014) found in their study that practitioners' attitudes towards research were the strongest predictor of research-based information use. Some educators are even sceptical about the overall value of educational research (e.g., Nägel et al., 2023; Gore & Gitlin, 2004; Nicholson-Goodman & Garman, 2007; Vanderlinde & van Braak, 2010), an idea that has been described in the context of science scepticism.

### *Science scepticism towards educational research*

While science scepticism is acknowledged as a phenomenon in various disciplines and topics, such as climate change or the COVID pandemic, to our best knowledge there are no prior studies on science scepticism towards science education research. However, there is research on a potential general scepticism of teachers towards broader educational research (e.g., Nägel et al., 2023; Nicholson-Goodman & Garman, 2007; Vanderlinde & van Braak, 2010). A study examining the personal and structural factors influencing evidence-based practices of teachers indicates that the relevance of current research findings to teaching practice is perceived as low by teachers (Ackeren et al., 2013). Educators' sceptical attitudes towards educational research range from criticizing that researchers fail to address questions relevant to practice (Vanderlinde & van Braak, 2010) to more cynical perspectives and distrust towards scientific research in general (Nicholson-Goodman & Garman, 2007). Additionally, science teachers are sometimes critical towards teaching material developed based on scientific evidence (Breuer, 2021); regarding the implementation of knowledge into practice, teachers even tend to perceive the knowledge of experienced colleagues as more trustworthy than scientific evidence (Ackeren et al., 2013). Although science-related beliefs are highly important (Joram et al., 2020; Schaik et al., 2018) for the reception of scientific knowledge, Schmidt et al. (2022) argue that the outcome of many studies indicating that teachers often refuse to utilize scientific evidence for their everyday practice is not caused by a general scepticism towards science's ability to provide stable and trustworthy knowledge. Instead, it might be due to the association of scientific evidence with more abstract and theoretical information (e.g., Bråten & Ferguson, 2015; Buehl & Fives, 2009; Ophoff & Cramer, 2022). Schmidt et al. (2022) explain this phenomenon with reference to confirmation bias: Even if teachers do not exhibit a general distrust of scientific evidence, confirmation bias functions as a filter in the evaluation of such evidence. It prevents teachers from changing their practice based on scientific findings when these do not align with their existing beliefs. Moreover, teachers believe that researchers cannot contribute to solving their educational problems because these problems are too complex and researchers do not know what happens in a classroom (Broekkamp & Van Hout-Wolter, 2007; Lysenko et al., 2014). Thus, even if teachers do not generally doubt that science education research can provide trustworthy knowledge, they associate scientific evidence so strongly with abstract information that they refuse to consider this evidence for their practice at all. The focal point for scepticism, therefore, lies in doubting the practical value of science education research (cf. Nägel et al., 2023). In our article, this attitude is consistent with our understanding of science scepticism, which we define as an attitudinal concept. According to Nägel et al. (2023), science scepticism towards educational research primarily refers to scepticism about its practical value. Similarly, we define science scepticism towards science education research as scepticism regarding any potential positive impact of science education research on teaching practice. For example, if someone is convinced that the methods employed in science education research lead to results incompatible with practical teaching, or if they believe that researchers in science education are too far removed from teaching practice to comprehend the 'real-world problems' that teachers face, or if they perceive the results as merely of academic interest without any other practical value, then this will be an indicator of science scepticism. While Nägel et al. (2023) explore scepticism towards the relevance of scientific content to teaching practice from a more general perspective, as there has been little research on this topic to date, our aim is to specifically look at science scepticism towards science education research (SSSE). Of course, science education research can be defined in many different ways; however, in our work, we adopt a broad definition of science education research as "the structures, processes, products, and individuals involved in the systematic development of knowledge" of science education following Broekkamp and van Hout-Wolters (2007, p. 205).

### *Beliefs, attitudes and behaviour*

Beliefs and attitudes are both constructs that lack clear and universally accepted definitions and are at times even used interchangeably (Pajares, 1992). Nevertheless, both constructs are considered highly relevant for understanding human behaviour (Nespor, 1987; Ajzen, 2005). Most commonly, attitudes are understood as evaluative judgments towards an object (Ajzen, 2005). For example, an object may be evaluated as pleasant or unpleasant (Ajzen & Fishbein, 1975). A widely used conceptualization of attitudes is the tripartite model, which comprises cognitive, affective, and behavioural intention components (Triandis, 1975). Within this model, beliefs are regarded as the cognitive component of attitudes. Eagly and Chaiken (1993) further emphasize that beliefs can, at least in principle, be verified or falsified based on external, objective criteria. Attitudes, by contrast, generally do not share this characteristic. In our understanding, science scepticism towards physics education research reflects an evaluation regarding its practical applicability. This evaluation is not an objective assessment, but rather a subjective perspective held by a teacher—and, thus, constitutes an attitude. This attitude is, of course, influenced by the teachers' beliefs about what physics education research entails and how it operates—conceptions that can, in principle, be externally examined. This perspective is consistent with contemporary models that link beliefs, attitudes, and behaviour, such as the Reasoned Action Approach (Fishbein & Ajzen, 2010). According to this model, behaviour is directly determined by intention. Intention, in turn, is shaped by attitudes, subjective norm, and perceived behavioural control, each of which is a function of underlying beliefs. However, little is known about how these beliefs, and the attitudes based upon them, are formed (Levin, 2015; Ajzen & Fishbein, 2005). Ajzen and Fishbein (2005) identify a wide range of potential background factors that may influence beliefs and, consequently, attitudes, including individual, social, and informational factors. They also emphasise that the influence of such background factors must be established empirically and should be investigated in a theory-driven manner (Ajzen & Fishbein, 2005). To our knowledge, there is no established theory concerning the determinants of teachers' attitudes, and the empirical evidence remains limited. Nevertheless, there is some indication that beliefs may be shaped by career-related variables such as teaching experience, teacher education, and engagement with professional literature (Levin, 2015; Erens, 2017), which we classify as individual factors in Ajzen and Fishbein's (2005) model. Moreover, teachers' epistemological beliefs are considered a key component within the model of professional competence proposed by Baumert and Kunter (2006). We therefore consider it appropriate, as a first step, to explore professional variables as factors potentially influencing teachers' attitudes.

*Research questions*

Based on our review of the literature, we conclude that the perspective of teachers on physics education research has just rarely been explored so far and could contribute to uncovering the obstacles to a sustainable implementation of science education research findings in teaching practice. Thus, this study primarily aims to explore potential scepticism towards science education research and the factors influencing its manifestations.

Three research questions guide the analysis:

RQ1: What are physics teachers' attitudes towards the value of physics education research for practical teaching?
RQ2: Which types of teachers, who exhibit scepticism towards the scientific discipline of physics education research, can be identified?
RQ3: Is there an association between professional variables and teachers' science scepticism towards physics education research?

While RQ1 will be explored using interview data, RQ2 and RQ3 will utilize quantitative data obtained from a survey study, which is informed by the results of the interviews. It is assumed that teachers may differ in their attitudes toward science education research, and that these attitudes are likely shaped by a variety of unknown factors. As a result, it is plausible that such attitudes do not form a homogeneous, unidimensional scale. A person-centred analytical approach aimed at identifying distinct types of teachers therefore appears appropriate. Possible factors that could influence the attitudes of physics teachers are not provided by previous research, however, they might include professional variables such as teaching experience, teacher education (traditionally educated and non-traditionally educated), school type (grammar school teachers and non-grammar school teachers), frequency of reading instructional journals, and engagement in research, as operationalized by the attainment of a doctoral degree. The teaching experience of teachers could potentially affect SSSE, as it has been found that in the field of educational research, more experienced teachers value research less than pre-service teachers (Gore & Gitlin, 2004). Furthermore, it is hypothesized that the education of teachers influences their level of SSSE, as traditionally educated teachers and teachers with alternative entry paths into the profession may have different levels of knowledge about science education research. While in Germany and Austria (where the study was conducted) traditionally educated teachers learn about science education content (pedagogical content knowledge) during their studies, it is possible that non-traditionally educated teachers have had little to no exposure to science education research upon entering the teaching profession. Having extensive knowledge in a specific field allows individuals to make more differentiated judgments about a specific knowledge domain and thus might lead to the development of more sophisticated beliefs about the relevance of this knowledge (Bromme et al., 2008), which could manifest in the science scepticism of teachers. Furthermore, Levin and He (2008) found that teachers attribute their pedagogical beliefs, among other factors, to coursework during teacher education, as well as to their experiences gained during practical phases in school. Based on this, we assume that beliefs about physics education research – and thus the attitude towards it – might likewise be shaped both by teacher education and by teaching experience. Consequently, we expect differences in teachers' attitudes towards physics education research depending on their teaching experience and the nature of their teacher education. Another relevant aspect is the well-known phenomenon that teachers enter the profession with certain preconceptions formed through their years of observing teachers as students themselves – a process referred to as the "apprenticeship of observation" (Lortie, 1975). Teacher education programmes have been shown to impact these preconceptions and can modify teachers' attitudes (Markic & Eilks, 2013, Levin & He, 2008). Teachers without formal teacher education, however, bypass this formative stage and enter the profession with their original preconceptions largely intact. We therefore assume that traditionally educated teachers and non-traditionally educated teachers may differ in their attitudes, which could possibly also lead to differences in their level of scepticism towards physics education research. Furthermore, it is conceivable that these two groups differ in their teaching ability beliefs and teacher enthusiasm (Lucksnat et al., 2022), two factors that can influence science scepticism (Nägel et al., 2023). To our knowledge, there is no empirical evidence suggesting that attitudes towards physics education research differ by school type. However, in our exploratory interview study, it was suggested that physics education research tends to be more relevant for grammar schools and perceived as less important for non-grammar schools. Furthermore, Rüger and Scheer (2025) identified differences in attitudes towards evidence-based practice between general and special education teachers. Based on these findings, we assume that there may be differences in teachers' science scepticism towards physics education research depending on school type (grammar school vs. non-grammar school). It is also assumed that teachers with a doctoral degree may have developed more sophisticated beliefs about research and, therefore, are less sceptical towards it compared to teachers without a doctoral degree. In general, a PhD shows that an individual is familiar with research processes and a measure for engagement in research. Having been part of the research community might influence how a scientific discipline is seen. Lastly, it is supposed that the frequency of reading instructional journals (journals that communicate science education research findings to teachers such as, e.g., "The Physics Teacher") could influence SSSE, as the willingness to consider scientific findings from these journals for practical teaching presupposes a certain scientific affinity. Thus, physics teachers who read instructional journals more often might be less sceptical towards science education research.

**Methods**

*Participants and Design*

The explorative study consists of two parts. First, an interview study was conducted, followed by a survey study: The interviews aimed to explore the attitudes of physics teachers towards physics education research. Since the attitudes have not been researched in prior studies yet, a qualitative exploration potentially giving deep insights seemed to be the best approach. In the second step, the results of the interview study were used to design a survey study, allowing for an exploration of these attitudes in a larger sample of physics teachers.

The interview study included a sample of 13 in-service physics teachers from Germany, who had worked in the teaching profession for an average of $M = 12.5$ ($SD = 8.7$) years ranging from 0.25 to 24.5 years. To examine a broad spectrum of attitudes towards physics education research, a purposeful sample was selected for this study. This means that the following factors were considered when selecting participants: Teacher education, school type, attainment of doctoral degree. These factors are listed in Table 1.

**Table 1:** Participants' characteristics of the interview study sample.

| Professional variable | Value | N |
|---|---|---|
| Teacher education | Traditionally educated | 10 |
|  | Non-traditionally educated | 3 |
| School type | Grammar school | 6 |
|  | Non-grammar school | 7 |
| Doctoral degree | Doctoral degree | 4 |
|  | No doctoral degree | 11 |

*Note. N* = Count.

Based on these criteria, teachers were contacted via email from a pool of known teachers. It should be noted that those who agreed to participate in the interviews are likely to have a certain affinity with physics education research, which must be considered when interpreting the findings. Moreover, the purposeful sample is not representative, which may lead to certain biases. Nevertheless, we assume that the inclusion of teachers with diverse backgrounds has allowed us to capture a broad spectrum of attitudes towards physics education research. The following survey study was conducted with $N = 174$ physics teachers from Germany and Austria, with $N = 144$ traditionally educated (in an academic teacher education program including courses in science education) and $N = 30$ non-traditionally educated teachers (other academic qualification, e.g. a master's degree in pure physics including no formal science education courses), $N = 114$ teachers at grammar schools and $N = 60$ teachers at non-grammar schools, as well as $N = 36$ teachers with a PhD. The participants' average age was $M = 44.5$ ($SD = 11.7$) years and they had worked in their profession for an average of $M = 14.6$ ($SD = 10.8$) years ranging from 0 to 46 years. Of the 174 participating teachers, $N = 78$ were male, $N = 53$ female, none diverse, and $N = 43$ chose not to provide information about their gender.

*Measures*

*Step 1: Interviews*

For data collection, a guideline-based interview was utilized. This method is often employed for exploratory research projects since, in contrast to questionnaires, it allows for individual expressions that are necessary for reconstructing less-explored concepts. Given the lack of studies on this topic in science education, we regard this as an appropriate choice. The development of the interview guideline followed the SPSS procedure according to Kruse (2015, pp. 227–230), in which interview questions are first collected, then examined, and finally sorted and subsumed. In this study, possible interview questions related to overarching topics that could be significant for capturing attitudes towards physics education research were initially gathered. These topics include the use of physics education research findings in one's own teaching, the connection to physics education research during studies and in practice, as well as general questions about physics education research. Only afterwards, each question was examined to determine if it was suitable for answering the research question. The interview questions were then organized to allow for a thematically structured flow of the interviews. Additionally, possible follow-up questions were noted for cases where participants might not understand the question. Finally, the interview questions, along with any necessary follow-ups, were summarized to form the interview guideline. The interview guideline was piloted with three students who are pursuing a master's degree in physics education and have gained some practical experience. The pilot testing aimed to verify the clarity of all interview questions, which could be largely confirmed. Some formulations were optimized, but fundamental changes to the interview guidelines were not necessary.

*Step 2: Survey*

To examine the previously found attitudes towards physics education research in a larger sample, a total of 22 survey items were created from the attitudes found in the interview study. The items cover the seven main attitudes identified in the interviews, along with the various reasons teachers provided in the interviews to justify them. The survey study was conducted in two iterations: In the first iteration, participants were asked to indicate their level of agreement using a four-point Likert scale ranging from *Strongly Agree* to *Strongly Disagree*. In the second iteration, our items were integrated into a larger questionnaire, for which a six-point Likert scale had been agreed upon for reasons of consistency. Accordingly, we adapted our original four-point scale to align with this format. In comparison to the items for which high levels of agreement on the Likert scale indicate high science scepticism, those items where high agreement reflects a positive attitude towards physics education research were reverse-coded to ensure consistent interpretation of scale directionality. The data were consolidated and the Likert values normalized. Also, demographics to explore influencing factors on attitudes were collected, e.g., gender, teaching experience, teacher education (traditionally educated and non-traditionally educated), school type (grammar school teachers and non-grammar school teachers), frequency of reading instructional journals, and engagement in research, as operationalized by the attainment of a doctoral degree (cf. section "research questions").

**Analysis**

*Analysis of the interviews*

The interviews were recorded using a microphone and later transcribed in anonymized form. Consent for audio recording and storage of the interviews was obtained. On average, the interviews lasted 39 minutes, with a standard deviation of 13 minutes, the longest interview taking 69 minutes and the shortest around 25 minutes. After completing the interviews, the audio recordings were initially transcribed using the Sonix.ai software, resulting in a total of 185 pages of interview material. The transcripts were then processed according to the transcription rules outlined in Krüger and Riemeier (2014). The material was analysed using structuring content analysis (Kuckartz & Rädiker, 2022, p. 129) where the attitudes to be investigated were formed as categories from the interview material through an inductive process. To this end, all statements in the interview material that provided insights into attitudes towards physics education research were identified. To enable the assignment of these statements to the categories, a focused summary was formulated for each statement, aiming to identify and organize initially similar statements. Subsequently, categories representing attitudes were formed based on these summaries, and the statements were assigned to the categories. Due to our exploratory approach, the category system was developed inductively from the interview material by a single researcher. To ensure high coding quality, we followed the criteria outlined by Kuckartz and Rädiker (2024, p. 236f.) and applied additional measures to increase objectivity and reliability. Specifically, we adopted the consensual coding approach described by Hopf and Schmidt (1993) which is particularly suitable for high complex, exploratory category systems: The category system was applied to a specific portion of the interview material by a second coder, and the coding assignments of both coders were then compared to assess the level of agreement or discrepancies. In total, 18 out of 92 codings were reviewed, with 14 of these codings matching between both coders. In calculating Cohen's kappa, we followed the approach outlined by Kuckartz and Rädiker (2024); with 21 categories, the expected value of chance agreement is 1/21, resulting in a chance-corrected kappa of $\kappa = 0.77$ which can be considered as good (Altman, 1990). For the remaining four codings, a discussion was held to determine whether the coding of the first or second coder was more appropriate, and the assignments were adjusted if necessary.

*Analysis of the survey data*

Given the exploratory nature of this study and the absence of an established theory regarding the structure of attitudes towards physics education research, an exploratory factor analysis (EFA) was conducted on the 22 items to identify potential underlying structures in these attitudes. To address RQ2, latent profile analysis (LPA) was conducted. LPA aims to identify different types or profiles of individuals based on similar response patterns across specific variables and is an often-used method to identify "types" of individuals. It is, therefore, appropriate for RQ2. This method assumes that heterogeneous response behaviour originates from a mixture of $K$ subpopulations, or profiles, within which individuals exhibit similar response patterns (e.g., Spurk et al., 2020). Given the limited research on physics teachers' attitudes towards science education research and the absence of a priori known variables that explain the heterogeneous distribution of attitudes, the assumption of latent subpopulations and an exploratory approach appears justified (e.g., Berlin et al., 2014; Spurk et al., 2020). While our sample size of 174 is appropriate for conducting latent profile analysis, it should be noted that the reliability of the results is limited with smaller sample sizes (Sinha et al., 2021). We report *BIC* as statistical criteria for selecting class model, entropy, smallest posterior probability and size of the smallest class as classification diagnostics (Van Lissa et al., 2024; Weller et al., 2020). Finally, we report the p-value of the bootstrapped likelihood ratio test. LPA was conducted using the *tidyLPA* package in R (Rosenberg et al., 2018).

Usually, multinomial logistic regression is conducted to examine the relation between profile membership and external variables (following, e.g., Berlin et al., 2014; Pastor et al., 2007; Weller et al., 2020). We therefore adopt this method here to address RQ3. The methods are described more in detail in the findings section as some of the decisions for the statistical approach are data-driven.

**Results**

***RQ1: What are physics teachers' attitudes towards the value of physics education research for practical teaching?***

The analysis of the interviews resulted in a total of seven main categories and 19 subcategories that have been assigned to teachers' expressions during the interviews. The main categories (see Table 2) describe attitudes of physics teachers towards physics education research while the subcategories further specify these attitudes.

Teachers' Views on Physics Education Research

**Table 2:** Physics teachers' attitudes regarding physics education research, formed based on interview material using structuring content analysis.

| | Attitude | N |
|---|---|---|
| 1 | Physics education research is relevant to teaching practice. | 8 |
| 2 | Physics education research is very advanced. | 2 |
| 3 | Physics education research has intrinsic value. | 1 |
| 4 | Findings from physics education research play a minor role in lesson planning compared to teachers' own practical experience. | 11 |
| 5 | Physics education research findings cannot be implemented in teaching practice at all. | 11 |
| 6 | Physics education research and teaching practice do not cooperate well. | 9 |
| 7 | Engaging with physics education research findings for lesson planning is particularly necessary just in the first years after entering the profession. | 4 |

*Note.* $N$ = Number of teachers (of 13) that expressed the respective attitude at least once in the interviews. Attitudes 1-3 are in general positive, while 4-7 are in general negative attitudes.

Both sceptical (attitudes 4–7) and non-sceptical attitudes (attitudes 1–3) towards science education research emerged from the interviews. The two most frequently expressed attitudes were rather sceptical. Eleven of the 13 teachers stated at least once that research findings play a minor role compared to their own practical experience (attitude 4). The reasons given for this attitude included the view that research findings would limit teachers' autonomy in designing their lessons, the perceived greater value of personal experience in shaping effective teaching, and unwillingness to modify lesson plans once they have been worked out and tried.

Additionally, 11 of the 13 teachers expressed at least once that physics education research findings cannot be implemented into teaching practice at all (attitude 5). Reasons included a lack of time to engage with research, the perception that research findings are too theoretical, misaligned with curricula, or irrelevant to the needs of non-grammar schools.

Nine of the 13 teachers indicated that cooperation between researchers and practitioners does not function well (attitude 6). This was primarily attributed to the assumption that university researchers lack an understanding of what is truly important for improving classroom teaching. Moreover, teachers perceived the value of physics education research to be particularly high in the early years of their careers (attitude 7), but its perceived importance declined over time in comparison to their own practical experience. This attitude contains both sceptical and non-sceptical elements.

However, eight of the 13 teachers also expressed at least once that physics education research is important for teaching (attitude 1). Seven teachers reasoned that physics education research addresses specific challenges encountered in physics teaching. Furthermore, two teachers considered physics education research to be a particularly advanced form of educational research (attitude 2), as it has already produced relevant findings (often mentioned are students' misconceptions) and continues to address current issues, in contrast to educational research in other subjects. One teacher stated that scientific findings in physics education research have intrinsic value, independent of their direct applicability to teaching, as they may contribute to fundamental research (attitude 3). Notably, some teachers expressed contradictory attitudes during the interviews (therefore, the numbers in Table 2 do not sum to 13): While they acknowledged the importance of physics education research for practical teaching, they still held sceptical views about its relevance and applicability.

These attitudes were reformulated into Likert-scale items for the quantitative study (research questions 2 and 3) to explore them within a larger sample and to identify different teacher profiles based on their attitudes. This seems particularly important, as the interview study itself revealed a diversity of perspectives.

### RQ2: Which types of teachers, who exhibit scepticism towards the scientific discipline of physics education research, can be identified?

*Exploratory factor analysis*

An exploratory factor analysis was conducted with the 22 items formed from the results found in the interview study to identify potential structures within the variables. Since Bartlett's test is significant ($\chi^2(231) = 1317$, $p < 0.001$) and all 22 items have an *MSA*-value (Measure of sampling adequacy) above 0.5 (overall *MSA*-value = 0.823), the prerequisites for conducting an exploratory factor analysis are met (Bartlett, 1950; Dziuban & Shirkey, 1974; Kaiser & Rice, 1974). To determine the number of factors, a parallel analysis was conducted, as it provides more reliable results compared to the scree plot or Kaiser criterion (O'Connor, 2000; Velicer et al., 2000). The analysis identified four factors. Since it is reasonable to assume that attitudes involve some degree of correlation between factors, an oblique rotation method was applied (Costello & Osborne, 2005). Following the recommendation of Hair et al. (2010), factor loadings below 0.45 were suppressed in accordance with the sample size of $N = 174$. This resulted in a well-interpretable solution comprising 14 variables without cross-loadings. The remaining eight variables have factor loadings below 0.45 across all four factors and were therefore excluded. The factor interpretation, example items and descriptive statistics are presented in Table 3.

**Table 3:** Resulting factors from the exploratory factor analysis, sample items, and the Cronbach's α, means and standard deviations for the resulting scales (where 1 represents total agreement and 0 indicates no agreement with the statement).

| *Factor (Number of Variables)* | *Example item* | *α* | *M* | *SD* |
|---|---|---|---|---|
| F1: Relevance of science education research for the individual teacher (5) | F1.3: *I have no need to rely on physics education research findings because, through my teaching experience, I have formed my own perception of how to design effective physics lessons.* | 0.78 | 0.27 | 0.19 |
| F2: Feasibility of physics education research findings for teaching practice (4) | F2.4: *Physics education research findings cannot be implemented in teaching practice at all because they are too theoretical and lack relevance to school practice.* | 0.78 | 0.53 | 0.19 |
| F3: Relevance of science education research for teaching practice in general (2) | F3.1: *Physics education research plays a crucial role in teaching practice because it addresses specific challenges encountered in physics teaching.* | 0.75 | 0.40 | 0.22 |
| F4: Cooperation between science education research and school practice (3) | F4.2: *Physics education research and school practice do not cooperate well because science education research findings do not reach schools.* | 0.75 | 0.64 | 0.20 |

*Note.* $\alpha$ = Cronbach's α; *M* = Mean value; *SD* = Standard deviation.

*Latent profile analysis*

We tested solutions with up to five classes, as the size of the smallest class becomes very small with additional classes. We selected *BIC* as the criterion for model selection, as it appears to be the most accurate index for determining the best model, especially with continuous variables (Morgan, 2015; Nylund et al., 2007). Entropy summarized class separability and values above 0.8 are acceptable (Weller et al., 2020). The posterior probabilities represent the certainty of class assignment for each profile. Although there is no standard cutoff value, values above 0.7 are generally recommended (Masyn, 2013; Spurk et al., 2020). Other authors consider values above 0.9 to be ideal and values above 0.8 to be acceptable (Weller et al., 2020). Following Weller et al. (2020), we report the minimum posterior probability (*Prob$_{min}$*) to assess the classification certainty. Regarding the size of the smallest class (*N$_{min}$*), a minimum of 25 individuals is recommended (Lubke & Neale, 2006). The bootstrapped likelihood ratio test (*BLRT*) compares neighbouring models and provides a *p*-value. A significant *p*-value ($p < 0.05$) indicates that a model with *K* profiles is superior to a solution with *K-1* profiles (Nylund et al., 2007).

**Table 4:** Quality criteria for the tested classes.

| Classes | npar | BIC | Entropy | Prob$_{min}$ | N$_{min}$ | p$_{BLRT}$ |
|---|---|---|---|---|---|---|
| 1 | 15 | 232.11 | 1.00 | 1.00 | 174 | - |
| 2 | 30 | 0.83 | 0.91 | 0.91 | 32 | 0.01 |
| 3 | 45 | -91.48 | 0.90 | 0.92 | 26 | 0.01 |
| 4 | 60 | **-110.30** | 0.84 | 0.87 | 24 | 0.01 |
| 5 | 75 | -80.80 | 0.86 | 0.89 | 13 | 0.02 |

*Note.* **npar** = *Number of parameters;* **BIC** = Bayesian information criterion; **Prob$_{min}$** = Minimum posterior probability; **N$_{min}$** = Size of the smallest class; **p$_{BLRT}$** = *p*-value of the bootstrapped likelihood ratio test.

As indicator variables, all 14 variables identified as relevant in the EFA were included instead of the four factors. Wurpts and Geiser (2014) found that models including a greater number of indicator variables are generally advantageous. They improve class assignment accuracy and can compensate for small sample sizes. Moreover, the authors recommend avoiding models with fewer than five indicator variables and point out that a larger number of indicator variables increases the number of possible response patterns. Given our exploratory approach, it seemed reasonable to allow for the identification of less common response patterns that might have remained undetected if we had used the four factors as indicator variables. Table 4 presents the fit indices and classification diagnostics for models with up to 5 classes. The model with the smallest *BIC* value is the solution with 4 classes. Entropy and minimum posterior probability of this model indicate an acceptable class separability and classification certainty. The size of the smallest class, at 24, is slightly below the recommended threshold of 25; however, this is justified by the relatively small sample size ($N = 174$). Furthermore, since the smallest class size in the next-best model (3 classes) is only marginally higher (*N$_{min}$* = 26), we opted for the model with the lowest *BIC*. The *BLRT* is significant for all models, indicating that the 5-class solution is significantly better than the 4-class solution. However, the 5-class solution is not considered a viable option due to its very small smallest class size (*N$_{min}$* = 13) and a higher *BIC* compared to the 3- and 4-class solutions. In addition to the tendency of the *BLRT* to overestimate the number of classes (Morin & Marsh, 2014), further limitations of the *BLRT* are widely discussed in the literature (e.g., Sinha et al., 2021; Van Lissa et al., 2024). Additionally, similar to findings by Sinha et al. (2021), the *BLRT p*-value was significant for all our tested models (we tested models with up to 10 classes), which suggests limited utility for this metric in the current context. Finally, we decided the model with four classes to be the best solution. To better interpret the profiles, we calculated the mean values for each factor and then plotted the averages. Figure 1 shows a profile plot of the four profiles.

The four classes correspond with four different types of physics teachers regarding their SSSE. The four types differ in their judgement of the four factors of SSSE that have been identified in the exploratory factor analysis.

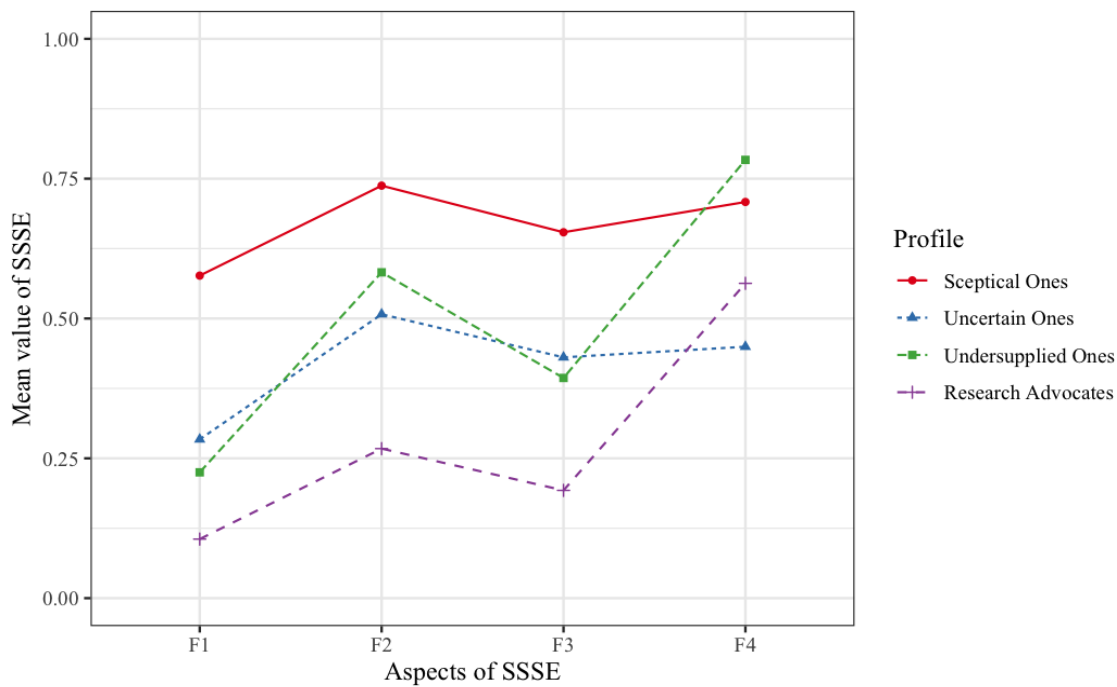

**Figure 1:** Profile plot of the four profiles. F1: Relevance of science education research for the individual teacher; F2: Feasibility of science education research findings for teaching practice; F3: Relevance of science education research for general teaching practice; F4: Cooperation between science education research and school practice.

*Profile 1: The Sceptical Ones (N = 24 (14%))*. The first profile is referred to as the *Sceptical Ones:* They score high in all areas of SSSE, indicating they attribute no value to science education research, neither for themselves personally nor for school practice in general. Additionally, they consider science education research findings to be impractical for classroom implementation and criticize the cooperation between research and schools.

*Profile 2: The Uncertain Ones (N = 47 (27%))*. The second profile shows a rather positive attitude regarding the relevance of physics education research for the individual teacher and nearly neutral values for all other factors. Teachers within this profile do attribute personal relevance to physics education research, yet they express neither a clear stance on its feasibility in the classroom nor on its relevance for general teaching practice. As a result, it remains unclear what this personal relevance actually entails for them. Moreover, in contrast to the other three profiles, this group does not exhibit scepticism regarding the cooperation between research and school practice but rather positions itself neutrally. This profile is the most difficult to interpret, as the values are generally close to the neutral point of 0.5, making clear conclusions impossible. Since this profile is neither clearly sceptical nor strongly aligned with research, it is referred to as *The Uncertain Ones*.

*Profile 3: The Undersupplied Ones (N = 71 (41%))*. The third profile is characterized by attributing relevance to physics education research both personally and in general while criticizing the feasibility of its findings. Additionally, this profile shows the highest scores for cooperation between research and schools among all profiles. This suggests that these teachers are even more critical of the research-practice cooperation than *The Sceptical Ones*. Considering the interview data, we interpret this profile as representing teachers who do attribute practical relevance to physics education research, yet perceive its current offerings as insufficiently aligned with their professional needs. One possible explanation for this perception is a lack of cooperation between physics education research and school practice from teachers' perspective, as suggested by the high values observed for this factor. From the perspective of this profile, physics education research could be more relevant to teaching practice and better able to supply teachers with findings if it had a better understanding of the needs present in school practice. For this reason, we have referred to this profile as *The Undersupplied Ones*.

*Profile 4: The Research Advocates (N = 32 (18%))*. The fourth profile has the lowest scores in almost all items indicating the lowest scepticism towards science education research. The *Research Advocates* attribute relevance to science education research both personally and in general and consider its findings to be practically implementable. Only the cooperation between research and school practice is not rated as positively by them as the other three aspects of SSSE.

### *RQ3: Is there an association between professional variables and teachers' science scepticism towards physics education research?*

Relations between profile membership and external variables were examined using multinomial logistic regression. This method allows for the simultaneous analysis of multiple variables with different levels of measurement within a single model and presents the results in a clear and concise manner. In contrast, conducting separate ANOVAs and Chi-squared tests for each variable with four groups (the four profiles) would lead to substantial inflation of the family-wise error rate due to the high number of comparisons. Multinomial logistic regression estimates how each predictor affects the odds of belonging to a particular group compared to a reference group, while controlling for all other variables in the model. In our model, profile membership was the dependent variable, with *The Sceptical Ones* as the reference group, and teaching experience, teacher education, school type, attainment of a doctoral degree, and frequency of reading instructional journals as covariates. The overall regression model is significant ($\chi^2(18) = 42.3$, $p = 0.001$) with a McFadden $R^2$ of $R^2 = 0.132$. Table 5 shows the results of the multinomial logistic regression model. Among all potential predictors, only the frequency of reading instructional journals has a significant effect on profile membership. The results suggest that reading instructional journals increases the likelihood of belonging to Profile 2 (*The Uncertain Ones*) ($\chi^2(1) = 3.42$, $p < 0.001$) and Profile 4 (*The Research Advocates*) ($\chi^2(1) = 2.87$, $p < 0.01$) (and, thus, reduce the likelihood to belong to Profile 1 (*Sceptical Ones*) and Profile 3 (*Undersupplied Ones*).

**Table 5:** Results of the multinomial logistic regression model.

|  | b | SD | z-value | p-value | Odds-ratio | 95% confidence interval |
|---|---|---|---|---|---|---|
| ***The Uncertain Ones* vs *The Sceptical Ones*** | | | | | | |
| Teaching experience | -0.022 | 0.031 | -0.73 | 0.47 | 0.98 | 0.92; 1.0 |
| Teacher education | -0.52 | 0.96 | -0.54 | 0.59 | 0.60 | 0.09; 3.9 |
| School type | -0.33 | 0.71 | -0.47 | 0.64 | 0.72 | 0.18; 2.9 |
| Doctoral degree | 1.2 | 0.98 | 1.2 | 0.22 | 3.3 | 0.48; 23 |
| Frequency of reading instructional journals | 2.4 | 0.61 | 3.9 | **<0.001** | 11 | 3.2; 35 |
| ***The Undersupplied Ones* vs *The Sceptical Ones*** | | | | | | |
| Teaching experience | -0.036 | 0.028 | -1.3 | 0.20 | 0.97 | 0.91; 1.0 |
| Teacher education | -0.11 | 0.82 | -0.13 | 0.90 | 0.90 | 0.18; 4.5 |
| School type | -0.37 | 0.62 | -0.60 | 0.55 | 0.69 | 0.20; 2.3 |
| Doctoral degree | 0.015 | 0.96 | 0.016 | 0.99 | 1.0 | 0.15; 6.7 |
| Frequency of reading instructional journals | 0.99 | 0.55 | 1.8 | 0.074 | 2.7 | 0.91; 7.9 |
| ***The Research Advocates* vs *The Sceptical Ones*** | | | | | | |
| Teaching experience | -0.046 | 0.034 | -1.3 | 0.18 | 0.96 | 0.89; 1.0 |
| Teacher education | -1.0 | 1.1 | -0.89 | 0.38 | 0.36 | 0.039; 3.4 |
| School type | 0.14 | 0.75 | 0.18 | 0.85 | 1.1 | 0.26; 5.0 |
| Doctoral degree | 0.84 | 1.1 | 0.77 | 0.44 | 2.3 | 0.28; 19 |
| Frequency of reading instructional journals | 2.4 | 0.65 | 3.8 | **<0.001** | 11 | 3.2; 40 |

*Note.* Reference group: *The Sceptical Ones*; Reference groups for categorical covariates: Teacher education: *traditional teacher education*; School type: *grammar school*; Doctoral degree: *no doctoral degree*.

**Discussion and implications**

Our study assumed that the effect that teachers tend to resist engaging with scientific evidence and rely more on their own experiences (e.g., Bråten & Ferguson, 2015; Buehl & Fives, 2009; Ophoff & Cramer, 2022; Landrum et al., 2002;), could possibly be related to a sceptical attitude of physics teachers towards science education research. In the exploratory factor analysis, we found four factors that contribute to SSSE: (1) Relevance of science education research for the individual teacher, (2) feasibility of physics education research findings for teaching practice, (3) relevance of science education research for teaching practice in general, and (4) cooperation between science education research and school practice.

We identified a science-sceptical profile ($N = 24$ (14%)), which, like all profiles, criticizes the cooperation between research and school practice but goes even further by questioning the value of science education research for school practice in general. This confirms indications of sceptical attitudes towards educational research in general: Vanderlinde and van Braak (2010) found that some teachers believe educational research has no inherent value for school practice, as it does not ask questions of practical relevance, and can therefore be ignored. Lysenko et al. (2014) also state that practitioners perceive their problems in schools as unsolvable by research, which likewise questions the inherent value of educational research.

The largest group we identified, *The Undersupplied Ones* ($N = 71$ (41%)), acknowledges the general value of science education research but sees its applicability in the classroom as limited. This profile exhibits high levels of scepticism regarding both the applicability of science education research findings and the cooperation between research and practice. Interestingly, *The Undersupplied Ones* report reading instructional journals more frequently than *The Sceptical Ones*. This suggests that although they engage with research by reading journals, they do not consider its findings to be implementable in practice, which could explain their critical stance towards the level of cooperation between research and school practice. Gore and Gitlin (2004) also identified teachers who acknowledge the value of educational research but lack the time to engage with its findings and implement them in their teaching practice. Consistent with our interview results, they found that research findings are not presented in a format that facilitates practical application, highlighting a broader critique of the cooperation between research and practice. In our interview study, we found that while many teachers acknowledge the value of research, they often emphasize that conditions vary across schools, making generalizations of research findings inadmissible. These varying conditions make the implementation of research findings nearly impossible (Joram et al., 2020).

Gore and Gitlin (2004) found that teachers stated that they were indeed interested in educational journals available in the teachers' lounge (e.g., Education Week), but simply lacked the time to read them. These findings are in line with our result that teaching journals are read least frequently by *The Sceptical Ones* and most frequently by the Research Advocates and that the frequency of reading teaching journals partially predicts profile membership. However, many personal variables – such as teaching experience, holding a PhD, or being a traditionally or alternatively qualified teacher (with the latter having had no exposure to science education research during their studies) – do not predict profile membership. Considering the difficulties involved in measuring the frequency of reading instructional journals (see Limitations), it follows that the potential predictor variables explored in this study may not adequately explain the identified profiles, if at all. However, frequency of reading instructional journals as the only significant predictor may provide an indication of other, yet unexplored, explanatory variables: In our theoretical framework, we have conceptualised the frequency of reading instructional journals as an individual background factor (Ajzen & Fishbein, 2005). However, in contrast to the other variables (such as teacher education, teaching experience, school type, or holding a PhD), this factor could also be understood as an informational factor (cf. Ajzen & Fishbein, 2005). This may indicate that informational factors relevant to educational contexts – such as content knowledge, pedagogical content knowledge, and general pedagogical knowledge (e.g., Shulman, 1987) – could play a role in predicting profile membership and should therefore be considered as potential background factors in future research.

Furthermore, we found that all profiles exhibited sceptical attitudes towards the cooperation between research and practice, which confirms existing findings. In many studies, including ours, teachers accuse researchers of being unaware of what happens in the classroom and of distancing themselves too much from 'reality' (e.g., Gore & Gitlin, 2004; Greenwood & Abbott, 2001). In an extreme view, this would imply that teachers think that researchers are entirely disconnected from school practice: Teachers in our interview study, as well as in the study by Gore and Gitlin (2004), referred to the so-called 'ivory tower' in which researchers reside, from where they are unable to gain insight into the real problems faced by teachers. This image of science education research and how science education researchers work suggests that teachers, with this perspective, question whether science education can fundamentally produce relevant outcomes at all, thus laying the groundwork for science scepticism towards science education research. Therefore, it would be important to examine in further studies what perceptions teachers have of how science education research works (a "nature of science education research"). It might even be plausible that the four factors are interrelated: For instance, teachers' criticism regarding the feasibility of findings from physics education research (Factor 2) may be rooted in a perceived lack of cooperation between schools and researchers (Factor 4). Moreover, the perceived lack of relevance – both for the individual teacher (Factor 1) and for general teaching practice (Factor 3) – may be attributable to the view that findings from physics education research are not practically applicable (Factor 2).

We argue that our study has important implications. First, physics teachers are not generally sceptical, and one could even suggest that they are slightly less sceptical towards physics education research than teachers in other studies are towards general educational research (Nägel et al., 2023). However, a notable degree of scepticism towards physics education research exists among physics teachers. Given the goal of evidence-based teaching to improve practice, this scepticism is concerning. Indeed, such scepticism towards science education (SSSE) may contribute to teachers' reluctance to use teaching materials developed through research. To address this issue, research should perhaps be more prominently integrated into physics teacher education, including professional development in practice. Teachers may currently lack awareness of how research in this field operates.

The second important implication is that most teachers are sceptical about the cooperation between research and practice (Factor 4). Factor 2 provides hints explaining this observation. For example, a key reason for this scepticism might be their perception that the research topics in physics education are too distant from their practical needs. This may indicate that research often overlooks the needs of practitioners or that the communication of research findings needs improvement. As researchers in the field, we might consider working more closely with teachers rather than focusing solely on what receives funding. Additionally, greater practical teaching experience for researchers—and, conversely, more research experience for teachers—could be beneficial. Time may be a limiting factor: Teachers report having no time to read research findings, and university researchers may have too little time to engage with schools or gain teaching experience in school settings. At the very least, further research is needed to identify effective ways to bridge the gap between research and practice. Renkl (2022) describes more ways on how research findings might be communicated effectively to teachers.

**Limitations**

The interview study addressing RQ1 is based on a small sample of 13 physics teachers. While the sample was purposefully selected, this still represents a limitation as potentially other attitudes might have been expressed by other teachers. Therefore, the results related to RQ1 should be viewed as an initial exploration of the topic and interpreted with caution. However, these findings informed the development of the questionnaire, which formed the quantitative core of the study and allows for greater generalizability. Nonetheless, this sample was also a convenience sample, meaning also this study remains exploratory—an appropriate approach given the current state of research. Physics teachers who hold a very negative attitude towards science education research might not have participated in the study in the first place. Consequently, the SSSE in the sample we examined would tend to be more positive than in the overall population of physics teachers. The internal validity of this study is limited by the Hawthorne effect: Participating teachers were aware that they were taking part in a science education study on teachers' attitudes towards physics education research. This awareness could have influenced the responding behaviour of the participating teachers. Additionally, LPA is a probabilistic approach in which each participant is assigned a certain probability for each of the four profiles (e.g., Berlin et al., 2014). For the multinomial logistic regression, we assigned each participant to the profile with the highest probability (so-called modal assignment (Pastor et al., 2007)), which disregards classification error (Van Lissa et al., 2024).

Furthermore, the results concerning the frequency of reading instructional journals should be interpreted with caution. Firstly, the assumption that the Likert scale values representing the frequency of reading instructional journals are interval-scaled is common, but just partially justified. Secondly, response patterns with the same mean can have different implications: For instance, an individual who reads a single journal very frequently might have the same mean score as someone who reads several journals occasionally, despite their reading habits differing significantly. Therefore, the mean provides only limited insight into the actual reading frequency of instructional journals. We recognize certain limitations in our multinomial logistic regression analyses. In particular, the relatively small number of cases in some profiles (e.g., $N = 24$) may affect the robustness and generalizability of the estimates. Moreover, while multicollinearity among predictors could influence the results, this was not specifically examined in the current study. These factors should be kept in mind when interpreting our findings.

**References**

Ackeren, van I., Binnewies, C., Clausen, M., Demski, D., Dormann, C., Koch, A. R., Laier, B., Preisendoerfer, P., Preuße, D., Rosenbusch, C., Schmidt, U., Stump, M., & Zlatkin-Troitschanskaia, O. (2013). Welche Wissensbestände nutzen Schulen im Kontext von Schulentwicklung? Theoretische Konzepte und erste Befunde des EviS-Verbundprojektes im Überblick [Which Bodies of Knowledge Do Schools Use in the Context of School Development? Theoretical Concepts and Initial Findings from the EviS Collaborative Project – An Overview]. In van I. Ackeren, M. Heinrich, & F. Thiel (Eds.), *Evidenzbasierte Steuerung im Bildungssystem? Befunde aus dem BMBF-SteBis-Verbund* (pp. 51–73). Waxmann.

Ajzen, I. (2005). *Attitudes, personality and behavior* (2nd Ed.). Open Univ. Press.

Ajzen, I., & Fishbein, M. (1975). A Bayesian analysis of attribution professes. *Psychological bulletin*, 261.

Altman, D. G. (1990). *Practical Statistics for Medical Research (Chapman & Hall / CRC Texts in Statistical Science)*. Taylor & Francis Ltd.

Altrichter, H., & Wiesinger, S. (2004). Der Beitrag der Innovationsforschung im Bildungssystem zum Implementierungsproblem [The Contribution of Innovation Research in the Education System to the Implementation Problem]. In *Psychologie des Wissensmanagements* (pp. 220–233).

Bartlett, M. S. (1950). TESTS OF SIGNIFICANCE IN FACTOR ANALYSIS. British Journal of Statistical Psychology, 3(2), 77–85. https://doi.org/10.1111/j.2044-8317.1950.tb00285.x

Baumert, J., & Kunter, M. (2006). Stichwort: Professionelle Kompetenz von Lehrkräften [Professional competence of teachers]. Zeitschrift für Erziehungswissenschaft, 9(4), 469–520. https://doi.org/10.1007/s11618-006-0165-2

Berlin, K. S., Williams, N. A., & Parra, G. R. (2014). An Introduction to Latent Variable Mixture Modeling (Part 1): Overview and Cross-Sectional Latent Class and Latent Profile Analyses. Journal of Pediatric Psychology, 39(2), 174–187. https://doi.org/10.1093/jpepsy/jst084

Bråten, I., & Ferguson, L. E. (2015). Beliefs about sources of knowledge predict motivation for learning in teacher education. Teaching and Teacher Education, 50, 13–23. https://doi.org/10.1016/j.tate.2015.04.003

Breuer, J. (2021). Implementierung fachdidaktischer Innovationen durch das Angebot materialgestützter Unterrichtskonzeptionen: Fallanalysen zum Nutzungsverhalten von Lehrkräften am Beispiel des Münchener Lehrgangs zur Quantenmechanik [Implementation of Subject-Specific Didactic Innovations Through the Provision of Material-Supported Teaching Concepts: Case Analyses of Teachers' Usage Behavior Using the Example of the Munich Course on Quantum Mechanics]. Logos Verlag.

Broekkamp, H., & van Hout-Wolters, B. (2007). The gap between educational research and practice: A literature review, symposium, and questionnaire. Educational

Research and Evaluation, 13(3), 203–220. https://doi.org/10.1080/13803610701626127

Bromme, R., Kienhues, D., & Stahl, E. (2008). Knowledge and Epistemological Beliefs: An Intimate but Complicate Relationship. In M. S. Khine (Ed.), Knowing, Knowledge and Beliefs (pp. 423–441). Springer Netherlands. https://doi.org/10.1007/978-1-4020-6596-5_20

Buehl, M. M., & Fives, H. (2009). Exploring Teachers' Beliefs About Teaching Knowledge: Where Does It Come From? Does It Change? The Journal of Experimental Education, 77(4), 367–408. https://doi.org/10.3200/JEXE.77.4.367-408

Costello, A. B., & Osborne, J. (2005). Best practices in exploratory factor analysis: Four recommendations for getting the most from your analysis. https://doi.org/10.7275/JYJ1-4868

Dicke, T., Holzberger, D., Kunina-Habenicht, O., Linninger, C., & Schulze-Stocker, F. (2016). „Doppelter Praxisschock" auf dem Weg ins Lehramt? Verlauf und potenzielle Einflussfaktoren emotionaler Erschöpfung während des Vorbereitungsdienstes und nach dem Berufseintritt ["Double Practice Shock" on the Path to Teaching? Trajectories and Potential Influencing Factors of Emotional Exhaustion During Teacher Training and After Entering the Profession]. Psychologie in Erziehung und Unterricht, 63(4), 244.

Dziuban, C. D., & Shirkey, E. C. (1974). When is a correlation matrix appropriate for factor analysis? Some decision rules. Psychological Bulletin, 81(6), 358–361. https://doi.org/10.1037/h0036316

Eagly, A. H., & Chaiken, S. (1993). The psychology of attitudes. Harcourt Brace Jovanovich College Publishers.

Edwards, A., Sebba, J., & Rickinson, M. (2007). Working with users: Some implications for educational research. British Educational Research Journal, 33(5), 647–661. https://doi.org/10.1080/01411920701582199

Erens, R. (2017). Entwicklung von Beliefs von Lehrkräften [Development of teachers' beliefs]. https://doi.org/10.17877/DE290R-18581

Fishbein, M., & Ajzen, I. (2010). Predicting and changing behavior: The reasoned action approach. Psychology Press.

Gore, J. M., & Gitlin, A. D. (2004). [RE]Visioning the academic–teacher divide: Power and knowledge in the educational community. Teachers and Teaching, 10(1), 35–58. https://doi.org/10.1080/13540600320000170918

Gräsel, C., & Parchmann, I. (2004). Implementationsforschung - oder: Der steinige Weg, Unterricht zu verändern [Implementation Research – Or: The Challenging Path of Changing Teaching Practices]. Unterrichtswissenschaft, 32(3), 196–214. https://doi.org/10.25656/01:5813

Greenwood, C. R., & Abbott, M. (2001). The Research to Practice Gap in Special Education. Teacher Education and Special Education: The Journal of the Teacher Education Division of the Council for Exceptional Children, 24(4), 276–289. https://doi.org/10.1177/088840640102400403

Hair Jr., J. F., Black, W. C., Babin, B. J., & Anderson, R. E. (2010). Multivariate Data Analysis (7th ed.).

Hopf, C., & Schmidt, C. (Eds.). (1993). Zum Verhältnis von innerfamilialen sozialen Erfahrungen, Persönlichkeitsentwicklung und politischen Orientierungen: Dokumentation und Erörterung des methodischen Vorgehens in einer Studie zu diesem Thema [On the Relationship Between Intra-Familial Social Experiences, Personality Development, and Political Orientations: Documentation and Discussion of the Methodological Approach in a Study on This Topic]. https://nbn-resolving.org/urn:nbn:de:0168-ssoar-456148

Joram, E., Gabriele, A. J., & Walton, K. (2020). What influences teachers' "buy-in" of research? Teachers' beliefs about the applicability of educational research to their practice. Teaching and Teacher Education, 88, 102980. https://doi.org/10.1016/j.tate.2019.102980

Kaiser, H. F., & Rice, J. (1974). Little Jiffy, Mark Iv. Educational and Psychological Measurement, 34(1), 111–117. https://doi.org/10.1177/001316447403400115

Krüger, D., & Riemeier, T. (2014). Die qualitative Inhaltsanalyse – eine Methode zur Auswertung von Interviews [Qualitative Content Analysis – A Method for Analysing Interviews]. In D. Krüger, I. Parchmann, & H. Schecker (Eds.), Methoden in der naturwissenschaftsdidaktischen Forschung (pp. 133–145). Springer Berlin Heidelberg. https://doi.org/10.1007/978-3-642-37827-0_11

Kruse, J. (2015). Qualitative Interviewforschung: Ein integrativer Ansatz [Qualitative Interview Research: An Integrative Approach] (2nd rev. ed.). Beltz Juventa.

Kuckartz, U., & Rädiker, S. (2022). Qualitative Inhaltsanalyse: Methoden, Praxis, Computerunterstützung: Grundlagentexte Methoden [Qualitative Content Analysis: Methods, Practice, Computer Support: Foundational Texts on Methods] (5th ed.). Beltz Juventa.

Kuckartz, U., & Rädiker, S. (2024). Qualitative Inhaltsanalyse. Methoden, Praxis, Umsetzung mit Software und künstlicher Intelligenz [Qualitative Content Analysis: Methods, Practice, Implementation using Software and Artificial Intelligence] (6th ed.). Juventa Verlag.

Landrum, T. J., Cook, B. G., Tankersley, M., & Fitzgerald, S. (2002). Teacher Perceptions of the Trustworthiness, Usability, and Accessibility of Information From Different Sources. Remedial and Special Education, 23(1), 42–48. https://doi.org/10.1177/074193250202300106

Levin, B. B. (2015). The Development of Teachers' Beliefs. In H. Fives & M. G. Gill (Eds.), International Handbook of Research on Teachers' Beliefs (pp. 48–65). Routledge.

Levin, B. & Ye He. (2008). Investigating the Content and Sources of Teacher Candidates' Personal Practical Theories (PPTs). Journal of Teacher Education, 59(1), 55–68. https://doi.org/10.1177/0022487107310749

Lorite, D. (1975). Schoolteacher: A Sociological Study. University of Chicago Press.

Lubke, G., & Neale, M. C. (2006). Distinguishing Between Latent Classes and Continuous Factors: Resolution by Maximum Likelihood? Multivariate Behavioral Research, 41(4), 499–532. https://doi.org/10.1207/s15327906mbr4104_4

Lucksnat, C., Richter, E., Schipolowski, S., Hoffmann, L., & Richter, D. (2022). How do traditionally and alternatively certified teachers differ? A comparison of their motives for teaching, their well-being, and their intention to stay in the profession. Teaching and Teacher Education, 117, 103784. https://doi.org/10.1016/j.tate.2022.103784

Lysenko, L. V., Abrami, P. C., Bernard, R. M., Dagenais, C., & Janosz, M. (2014). Educational Research in Educational Practice: Predictors of Use. Canadian Journal of Education, 37(2), 1–26.

Markic, S., & Eilks, I. (2013). POTENTIAL CHANGES IN PROSPECTIVE CHEMISTRY TEACHERS' BELIEFS ABOUT TEACHING AND LEARNING— A CROSS-LEVEL STUDY. International Journal of Science and Mathematics Education, 11(4), 979–998. https://doi.org/10.1007/s10763-013-9417-9

Masyn, K. E. (2013). Latent Class Analysis and Finite Mixture Modeling. In The Oxford Handbook of Quantitative Methods in Psychology (Vol. 2). Oxford University Press. https://doi.org/10.1093/oxfordhb/9780199934898.013.0025

Merzyn, G. (2004). Lehrerausbildung - Bilanz und Reformbedarf: Überblick über die Diskussion zur Gymnasiallehrerausbildung, basierend vor allem auf Stellungnahmen von Wissenschafts- und Bildungsgremien sowie auf Erfahrungen von Referendaren und Lehrern [Teacher Education – Assessment and Need for Reform: An Overview of the Discussion on Upper Secondary Teacher Education, Primarily Based on Statements from Academic and Educational Committees, as Well as the Experiences of Trainee Teachers and In-Service Teachers] (2nd rev. ed.). Schneider-Verl. Hohengehren.

Morgan, G. B. (2015). Mixed Mode Latent Class Analysis: An Examination of Fit Index Performance for Classification. Structural Equation Modeling: A Multidisciplinary Journal, 22(1), 76–86. https://doi.org/10.1080/10705511.2014.935751

Morin, A. J. S., & Marsh, H. W. (2014). Disentangling Shape from Level Effects in Person-Centered Analyses: An Illustration Based on University Teachers' Multidimensional Profiles of Effectiveness. Structural Equation Modeling: A Multidisciplinary Journal, 22(1), 39–59. https://doi.org/10.1080/10705511.2014.919825

Nägel, L., Bleck, V., & Lipowsky, F. (2023). "Research findings and daily teaching practice are worlds apart" – Predictors and consequences of scepticism toward the relevance of scientific content for teaching practice. Teaching and Teacher Education, 121, 103911. https://doi.org/10.1016/j.tate.2022.103911

Nelson, J., & Campbell, C. (2017). Evidence-informed practice in education: Meanings and applications. Educational Research, 59(2), 127–135. https://doi.org/10.1080/00131881.2017.1314115

Nespor, J. (1987). The role of beliefs in the practice of teaching. Journal of Curriculum Studies, 19(4), 317–328. https://doi.org/10.1080/0022027870190403

Nicholson–Goodman, J., & Garman, N. B. (2007). Mapping practitioner perceptions of 'It's research based': Scientific discourse, speech acts and the use and abuse of research. International Journal of Leadership in Education, 10(3), 283–299. https://doi.org/10.1080/13603120701257297

Niebert, K., & Gropengießer, H. (2014). Leitfadengestützte Interviews [Semi-structured Interviews]. In D. Krüger, H. Schecker, & I. Parchmann (Hrsg.), Methoden in der naturwissenschaftsdidaktischen Forschung (S. 121–132). Springer Berlin Heidelberg. https://doi.org/10.1007/978-3-642-37827-0_10

Nylund, K. L., Asparouhov, T., & Muthén, B. O. (2007). Deciding on the Number of Classes in Latent Class Analysis and Growth Mixture Modeling: A Monte Carlo Simulation Study. Structural Equation Modeling: A Multidisciplinary Journal, 14(4), 535–569. https://doi.org/10.1080/10705510701575396

O'Connor, B. P. (2000). SPSS and SAS programs for determining the number of components using parallel analysis and Velicer's MAP test. Behavior Research Methods, Instruments, & Computers, 32(3), 396–402. https://doi.org/10.3758/BF03200807

Ophoff, J. G., & Cramer, C. (2022). The Engagement of Teachers and School Leaders with Data, Evidence and Research in Germany. In C. Brown & J. R. Malin (Eds.), The Emerald Handbook of Evidence-Informed Practice in Education (pp. 175–195). Emerald Publishing Limited. https://doi.org/10.1108/978-1-80043-141-620221026

Pajares, M. F. (1992). Teachers' Beliefs and Educational Research: Cleaning Up a Messy Construct. Review of Educational Research, 62(3), 307–332. https://doi.org/10.3102/00346543062003307

Pastor, D. A., Barron, K. E., Miller, B. J., & Davis, S. L. (2007). A latent profile analysis of college students' achievement goal orientation. Contemporary Educational Psychology, 32(1), 8–47. https://doi.org/10.1016/j.cedpsych.2006.10.003

Renkl, A. (1996). Träges Wissen: Wenn Erlerntes nicht genutzt wird [Inert Knowledge: When What Is Learned Is Not Used]. Psychologische Rundschau, 47(2), 78–92.

Renkl, A. (2022). Meta-analyses as a privileged information source for informing teachers' practice?: A plea for theories as primus inter pares. Zeitschrift für Pädagogische Psychologie, 36(4), 217–231. https://doi.org/10.1024/1010-0652/a000345

Rosenberg, J., Beymer, P., Anderson, D., Van Lissa, C. j., & Schmidt, J. (2018). tidyLPA: An R Package to Easily Carry Out Latent Profile Analysis (LPA) Using Open-Source or Commercial Software. Journal of Open Source Software, 3(30), 978. https://doi.org/10.21105/joss.00978

Rüger, L., & Scheer, D. (2025). General and special education teachers' attitudes towards evidence-based practice. Journal of Research in Special Educational Needs, 25(2), 323–340. https://doi.org/10.1111/1471-3802.12727

Schaik, P. v., Volman, M., Admiraal, W., & Schenke, W. (2018). Barriers and conditions for teachers' utilisation of academic knowledge. International Journal of Educational Research, 90, 50–63. https://doi.org/10.1016/j.ijer.2018.05.003

Schmidt, K., Rosman, T., Cramer, C., Besa, K.-S., & Merk, S. (2022). Teachers trust educational science—Especially if it confirms their beliefs. Frontiers in Education, 7, 976556. https://doi.org/10.3389/feduc.2022.976556

Schrader, J., Hasselhorn, M., Hetfleisch, P., & Goeze, A. (2020). Stichwortbeitrag Implementationsforschung: Wie Wissenschaft zu Verbesserungen im Bildungssystem beitragen kann [Key Contribution on Implementation Research: How Science Can Contribute to Improvements in the Education System]. Zeitschrift für Erziehungswissenschaft, 23(1), 9–59. https://doi.org/10.1007/s11618-020-00927-z

Shulman, L. (1987). Knowledge and teaching: Foundations of the new reform. Harvard Education Review, 57(1), 1–22.

Sinha, P., Calfee, C. S., & Delucchi, K. L. (2021). Practitioner's Guide to Latent Class Analysis: Methodological Considerations and Common Pitfalls. Critical Care Medicine, 49(1), e63–e79. https://doi.org/10.1097/CCM.0000000000004710

Spurk, D., Hirschi, A., Wang, M., Valero, D., & Kauffeld, S. (2020). Latent profile analysis: A review and "how to" guide of its application within vocational behavior research. Journal of Vocational Behavior, 120, 103445. https://doi.org/10.1016/j.jvb.2020.103445

Stark, R., Herzmann, P., & Krause, U.-M. (2010). Effekte integrierter Lernumgebungen—Vergleich problembasierter und instruktionsorientierter Seminarkonzeptionen in der Lehrerbildung [Effects of Integrated Learning Environments – A Comparison of Problem-Based and Instruction-Oriented Seminar Concepts in Teacher Education]. Zeitschrift für Pädagogik, 56(4), 548–563. https://doi.org/10.25656/01:7159

Triandis, H. C. (1975). Einstellungen und Einstellungsänderungen [Attitudes and Attitude Changes]. Beltz Verlag.

Van Lissa, C. J., Garnier-Villarreal, M., & Anadria, D. (2024). Recommended Practices in Latent Class Analysis Using the Open-Source R-Package tidySEM. Structural Equation Modeling: A Multidisciplinary Journal, 31(3), 526–534. https://doi.org/10.1080/10705511.2023.2250920

Vanderlinde, R., & van Braak, J. (2010). The gap between educational research and practice: Views of teachers, school leaders, intermediaries and researchers. British Educational Research Journal, 36(2), 299–316. https://doi.org/10.1080/01411920902919257

Velicer, W. F., Eaton, C. A., & Fava, J. L. (2000). Construct Explication through Factor or Component Analysis: A Review and Evaluation of Alternative Procedures for Determining the Number of Factors or Components. In R. D. Goffin & E. Helmes (Eds.), Problems and Solutions in Human Assessment (pp. 41–71). Springer US. https://doi.org/10.1007/978-1-4615-4397-8_3

Vorholzer, A., Petermann, V., Weber, J., Zu Belzen, A. U., & Tiemann, R. (2024). Explicit instruction on procedural and epistemic knowledge – is it happening? A video-based exploration of classroom practice. Research in Science & Technological Education, 42(3), 513–532. https://doi.org/10.1080/02635143.2022.2153245

Weller, B. E., Bowen, N. K., & Faubert, S. J. (2020). Latent Class Analysis: A Guide to Best Practice. Journal of Black Psychology, 46(4), 287–311. https://doi.org/10.1177/0095798420930932

Wilhelm, T., Schecker, H., & Hopf, M. (Eds.). (2021). Unterrichtskonzeptionen für den Physikunterricht: Ein Lehrbuch für Studium, Referendariat und Unterrichtspraxis [Instructional Concepts for Physics Teaching: A Textbook for University Studies, Teacher Preparation, and Classroom Practice]. Springer Berlin Heidelberg. https://doi.org/10.1007/978-3-662-63053-2

Wurpts, I. C., & Geiser, C. (2014). Is adding more indicators to a latent class analysis beneficial or detrimental? Results of a Monte-Carlo study. Frontiers in Psychology, 5. https://doi.org/10.3389/fpsyg.2014.00920